\documentclass[twocolumn,prc,nofootinbib,showpacs]{revtex4}
\usepackage{graphicx}
\newcommand{\bdec}{$\beta$-decay }

\newcommand{\tehdz}{$^{123}$Te }
\newcommand{\sbhdz}{$^{123}$Sb }
\def\thepage{\roman{page}}
\def\be{\begin{equation}}
\def\ee{\end{equation}}
\begin{document}


\title{An alternative search for the electron capture of
\tehdz }
\author{D. M\"unstermann}
\author{K. Zuber}
\altaffiliation{Present address: Denys Wilkinson Laboratory,
University of Oxford, Keble Road, Oxford OX1 3RH}
\affiliation{Lehrstuhl f\"ur Experimentelle Physik IV,
Universit\"at Dortmund,\\ Otto--Hahn--Strasse 4, 44227 Dortmund,
Germany}

\begin{abstract}
A search for the second forbidden electron capture of \tehdz has
been performed.
A new technique for searches of rare nuclear decays using CdZnTe detectors
has been established. After a measuring time of 195 h 
no signal could be found resulting in a lower half-life limit of
$T_{1/2} > 3.2 \cdot 10^{16}$ yrs (95 \% CL) for this process.
This clearly discriminates between existing experimental results
which differ by six orders of magnitude and our data are in strong
favour of the result with longer half-lives.
\end{abstract}


\maketitle

\section{Introduction}
In the past decades the investigation of \bdec{} has played a
major role in understanding weak interactions and their structure.
Even though being a little bit out of fashion, there are still
interesting problems to be investigated. Among
them is the second forbidden unique electron capture (EC) of
\tehdz occuring to the ground state of \sbhdz with a transition
energy of $53.3 \pm 0.2$ keV \cite{a}.
Two positive evidences obtained for this decay
mode show a discrepancy by six orders of magnitude, namely $(1.24
\pm 0.10) \cdot 10^{13}$ yrs \cite{b}
and $(2.4 \pm 0.9) \cdot 10^{19}$ yrs \cite{c}.
Even with the lower value already in slight disagreement with other
bounds on the lifetime of K shell capture in \tehdz \cite{d}
it is quoted quite often in literature \cite{e}. However, still only a lower
bound of about $10^{13}$ yrs is used \cite{toi}.
One of the main problems of the past measurements was associated
with the fact that observations relied on the outside detection of
K X-rays following the deexcitation of $^{123}$Sb. This line is at
26.1 keV, very close to the Te K X-ray line at 27.3 keV
which might be excited by other
processes like cosmic rays and radioactive background.\\
In the measurement reported here, a new technique for searches of
rare nuclear decays using CdZnTe semiconductor detectors
has been used for the first time. These offers several advantages. The natural
abundance of \tehdz is 0.9 \%, hence it is intrinsic to any detector
containing Te. Due to this fact, the total cascade after K-capture
is contained within the detector resulting in a peak at 30.5 keV.
Cd(Zn)Te detectors are well known devices in
X-ray and $\gamma$-ray astronomy and offer good energy resolution.
The study was performed within a pilot project for the
COBRA-experiment \cite{COBRA}, which plans to use a large amount
of CdZnTe detectors to explore double beta decay.

\section{Experimental setup}

\begin{figure}
\includegraphics[width=8cm]{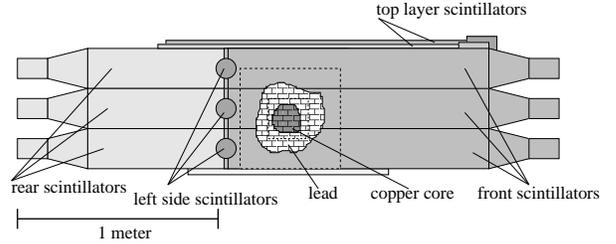}
\caption{\label{fig:setup}Schematic drawing of the setup used to perform
the measurement described.  The detector is surrounded by a passive shield of clean copper and lead
and an active shield against cosmic ray muons using plastic scintillators.  For details see text.}
\end{figure}

The measure\-ments were performed with a $10 \times 10 \times 5$
mm$^3$ CdZnTe detector provided by
eV-Products\footnote{eV-PRODUCTS, Saxonburg, PA, USA,
www.evproducts.com}. The detector is encapsulated within an
aluminium-tube. It is working with the coplanar grid technology,
focussing on the readout of electrons \cite{CPG}. This avoids
asymmetric peaks normally seen because of the insufficient
collection of holes due to trapping within CdZnTe.

The energy resolution was measured with various sources, among
them $^{241}$Am and $^{57}$Co, resulting in lines at 59.5, 122.1
and 136.5 keV respectively. An FWHM-energy resolution of $\Delta E
= 4.7$ keV is determined for 30.5 keV by extrapolation.\\
To be sensitive for rare decays, a special shielding had to be
constructed to reduce background events due to environmental
radioactivity and cosmic rays.
A schematic plot of the setup is shown in Fig.1.
In this way, a $50 \times 50 \times
50$ cm box made of copper and lead was designed. The inner $20
\times 20 \times 20$ cm consisted of electrolytic copper, the
outer part was made of spectroscopy lead. In a first step, the
copper was cleaned in an ultrasonic bath. All lead bricks were
etched shortly before installation with HCl and HNO$_3$ to clean
the surfaces, the copper was electropolished. The complete
shielding was surrounded by an air-tight box of aluminium, which
was flushed with nitrogen. This prevents air and with it
$^{222}$Rn from penetrating into the apparatus. To suppress
signals due to cosmic ray muons, an active veto consisting of 19
plastic scintillators was used. They covered all four sides of the
box in a single layer with about 95 \% detection efficiency as well
as the top side in two layers resulting in about 99.5 \%
efficiency.
All scintillators were fed into discriminators and a
common OR was used to form the veto signal. In that way after the
firing of any of the scintillators the data aquisition was blocked
for 20 $\mu$s
resulting in a dead-time of about $1.6$ \%. Only accepted signals
were fed into a multichannel analyser card in a PC, serving as
DAQ-system. The shielding depth was about 5 meter water equivalent (mwe).

\begin{figure} \includegraphics[width=8cm]{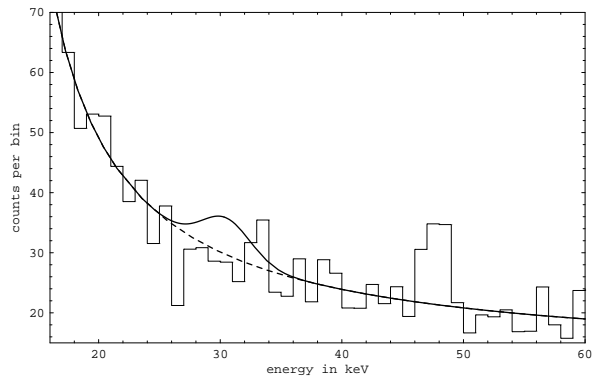}
\caption{\label{fig:spectrum} Measured spectrum in the region of
interest. No obvious peak is visible. The bin width is 1 keV.
Shown are the fit assuming a pure background distribution with an
$(1/E + const)$-dependence (dashed line) and the background fit
plus the excluded Gaussian peak (solid line).}
\end{figure}

\section{Results}
Single runs were taken with 30 minutes life-time each.
392 runs were obtained of which 2 had to be removed resulting in 
195 hours of good
data. The obtained spectrum of interest is shown in
Fig.2. No signal is visible at 30.5 keV. Fitting the region from
$17.5$ to $45.5$ keV with a $1/E + const$ function results in a
$\chi^2/\mbox{dof}$ of $11.7/11$ within the $2 \Delta$E-region
around the peak position, a fit including a Gaussian peak at the
right position and with the correct energy resolution gives a
$\chi^2/\mbox{dof}$ of $11.8/11$ while the gaussian content ist
consistent with zero. Excluding the $2 \Delta$E-region around
$30.5$ keV (i.e. 25 to 36 keV) from the fitpoints, the resulting
$\chi^2/\mbox{dof}$ of the pure background fit is still $12.0/11$.
Therefore, the background-only hypothesis is well within allowed
parameters.

Using the fit without the peak region and adding a Gaussian 
more than 31.2 events can be ruled out (95 \% CL). Knowing
the mass of $2.89$ g the upper bound can be converted into a lower
limit on the half-life for the electron capture of \tehdz: \be
T_{1/2} > 3.2 \cdot 10^{16} \quad \text{yrs} \quad 95 \% \text{CL}
\ee This clearly rules out possible observations with half-lives
of the order of $10^{13}$ yrs and is in strong favor of the
results of Alessandrello et al. \cite{c}. Longer half-lives are
also supported from theoretical considerations \cite{f,g}.

\section{Summary}
A novel technique for studying rare nuclear decays using CdZnTe semiconductor 
detectors was used to
investigate the second forbidden unique electron capture of
\tehdz. It takes advantage of the fact that source and detector
are identical. The main motivation for this search was to
discriminate between two published results which differ by six
orders of magnitude. No evidence for this process was found and a
half-life limit larger than $3.2 \cdot 10^{16}$ yrs (95 \% CL) is
obtained. This clearly rules out the solutions obtaining $10^{13}$
yrs. The measurement was performed during tests for the planned
COBRA-experiment to search for double beta decay. A proof for the
quoted half-life of $2.4 \pm 0.9 \cdot 10^{19}$ yrs can only be
done with a setup using a larger amount
of CdZnTe detectors.

\section{Acknowledgements}
We thank H. Kiel and C. G\"o{\ss}ling for
valuable discussions and support. We also acknowledge the support
of the mechanical
workshop of the University of Dortmund during the construction of
the test-setup. Finally we want to thank eV-Products for
generously lending the detector that was used for our
measurements.

\end{document}